\documentclass[5p]{elsarticle}

\usepackage[latin1]{inputenc}
\usepackage{amssymb}
\usepackage{lipsum}
\usepackage{amsthm}
\usepackage{amsmath}
\usepackage{slashed}

\journal{Physics Letters B}

\begin{document}

\begin{frontmatter}



\title{Extra dip in ultrahigh energy neutrino spectrum from generalized uncertainty principle}

\author[DCI]{J. Barranco}
%\email{jbarranc@fisica.ugto.mx}

\author[DCI]{Emiliano Dur\'an}
\affiliation[DCI]{organization={Departamento de F\'isica}, 
\addressline = {DCI, Campus Le\'on, Universidad de Guanajuato}, 
\city = {Le\'on},
\postcode = {37150},
\state = {Guanajuato}, 
\country = {M\'exico}}

\begin{abstract}
We revisited the scenario of a resonant enhancement in the oscillation probability due to an interaction potential between neutrinos and dark matter with the novelty of the inclusion of the generalized uncertainty principle. 
It is shown that a new resonant conversion appears at higher energies. This effect could be tested with future neutrino data as new dips in the ultrahigh energy neutrino flux.
\end{abstract}
\end{frontmatter}

\section{Introduction}

Neutrino oscillation is a pure quantum mechanical effect. First proposed by Pontecorvo \cite{Pontecorvo:1957cp,Pontecorvo:1957qd,Pontecorvo:1967fh,Bilenky:1978nj}, the neutrino oscillations are the effect where the flavor of the neutrino changes while traveling through space. This phenomenon was later confirmed by several atmospheric, solar and reactor neutrino experiments \cite{Super-Kamiokande:2010orq,SNO:2009uok,KamLAND:2010fvi,K2K:2006yov,DayaBay:2012fng}.
For neutrinos propagating through a material medium, they experience a resonance that enhances the oscillation probability. This enhancement in the neutrino flavor conversion is called the Mikheyev-Smirnov-Wolfenstein (MSW) effect \cite{Mikheev:1986wj,Wolfenstein:1977ue}. 
The MSW effect could appear not only in the propagation of neutrinos interacting with baryonic matter through the weak force. It is possible, for instance, that ultrahigh energy (UHE) active neutrinos mixed with sterile neutrinos could operate a MSW effect due to the interaction of neutrinos with dark matter as they travel within the galactic halo, as proposed in \cite{Miranda:2013wla}.
In this work we will re-consider the scenario proposed in \cite{Miranda:2013wla} with the inclusion of a generalized uncertainty principle (GUP) \cite{Maggiore:1993rv,Garay:1994en}. 
We will show that GUP induces a second resonance in the oscillation probability that could lead to observable dips in the UHE neutrino spectrum. 
Previous works that include GUP with neutrino oscillations phenomenology  \cite{Sprenger:2010dg,Torri:2024jwc,Gialamas:2024ivq} have not consider UHE neutrinos neither sterile neutrinos or DM interactions. Thus, the scenario proposed here could be of interest in those three research areas of neutrino physics. Furthermore, this work shows a possible connection with quantum gravity phenomenology since GUP could be seen as a generic feature of quantum gravity.  
Indeed, perhaps the best messengers that could shed light of quantum gravity effect on the highest energies  are the neutrinos because the Universe is impenetrable to photons with energies larger that $10^{15}$ eV$=1$ PeV. 

On the experimental side, it is worthy to recall that IceCube has observed the flux of electron and $\tau$ UHE neutrinos in the energy range from $16$ TeV to $2.6$ PeV \cite{IceCube:2020acn}. IceCube had also detected the muon neutrino flux \cite{Abbasi:2021qfz} and the detection  of a cascade of high-energy particle shower that is consistent with being created at the Glashow resonance \cite{IceCube:2021rpz}, i.e. IceCube has evidence of an electron antineutrino with energy of $6.3$ PeV. 
In particular, in the electron-muon neutrino spectrum, there is a hint of a deficit in the 200 TeV - 1 PeV energy range known as dip in ultra-high energy (UHE) neutrino flux.
Diverse explanations has been proposed to explain such dip \cite{Kamada:2015era,Chauhan:2021ixn,Bustamante:2020mep}.  
In this work we will show that if the observed deficit in the UHE neutrino spectrum is the result of a resonant conversion of UHE active neutrinos into sterile neutrinos, then, there should be a second dip in the spectrum due to the inclusion of a GUP in neutrino oscillations.
The paper is organized as follows: in section \ref{seccion2} we will derive the survival neutrino probability of active neutrinos to sterile neutrinos interacting in a dark matter medium with constant density including GUP. In section \ref{seccion3} we will derive the resonance condition and show that there is a second resonance induced by the GUP. We will draw some conclusions in section \ref{seccion4}.

\section{Neutrino oscillations with a generalized uncertainty principle}
\label{seccion2}
It is commonly suggested that in quantum gravity and in string theory there exist a minimum observable length \cite{Jaekel:1993ft,Konishi:1989wk}. This minimal length implies a minimal uncertainty in a position measurement giving rise to a modification of the Heisenberg algebra which serves as a foundation for a GUP \cite{Kempf:1994su}. This modified Heisenberg algebra can be expressed as a new commutation relation:
\begin{equation}
[x_i,p_j]=i\hbar\left(\delta_{ij}+\beta\delta_{ij}p^2+2\beta p_ip_j\right)\,,
\end{equation}
that can be expressed as a canonical commutation relation $[\Tilde{x_i},\Tilde{p_j}]=i\hbar$ for the variables with tilde \cite{Gialamas:2024ivq}
\begin{eqnarray}
x_i&=&\Tilde{x_i}\,,\nonumber \\
p_i&=&\Tilde{p_i}(1+\beta\Tilde{p_i}^2)\,,\nonumber\\
p_0&=&\Tilde{p_0}=E/c\,,
\end{eqnarray}
which also satisfies the canonical dispersion relation 
$\Tilde{p_0}^2-\Tilde{p_i}^2=m^2$. 
In terms of this new variables, the squared 4-momentum of the physical quantities can be expressed as
\begin{equation}
p_0^2-p_i^2=\Tilde{p_0}^2-\tilde{p}^2\left(1+2\beta\Tilde{p}^2+\mathcal{O}(\beta^2)\right)\,,
\end{equation}
that can be rewritten as
\begin{equation}
E^2=m^2 c^4+p^2 c^2(1-2\beta p^2 c^2)\,.\label{dispersion}
\end{equation}
With this new dispersion relation we can start our analysis of neutrino oscillation. From now on we will use natural units where $c=\hbar=1$.
Assuming that the neutrino masses $m$ are very small 
$m\ll p$, the dispersion relation Eq. (\ref{dispersion}) can be expanded as
\begin{equation}
E\simeq p\sqrt{1-2\beta p^2}+\frac{m^2}{2p\sqrt{1-2\beta p^2}}+\mathcal{O}(m^4)\,.
\end{equation}

Following \cite{Miranda:2013wla}, we consider the simplified scenario of an active neutrino $\nu_\alpha$ and one sterile neutrino $\nu_s$ interacting with a dark matter potential. The evolution equation will be
\begin{equation}
i\hbar \frac{\rm d}{{\rm d}t}
\left(\begin{array}{c}
\nu_\alpha \\
\nu_s \\
\end{array} \right)
=
{\rm H}_{eff}
\left( \begin{array}{c}
\nu_\alpha \\
\nu_{s} \\ \end{array} \right) \,.
\label{eq:1}
\end{equation}
Here, the effective Hamiltonian $\rm{H}_{eff}$ is given by
\begin{equation}
{\rm H}_{eff}=\frac{1}{2E\sqrt{1-2\beta E^2}}U^\dagger\left(\begin{array}{cc} 
m_\alpha^2 & 
0 \\ 
& \\ 
0 & 
m_s^2 \\
 \end{array} \right)U+V_{int} \, ,
\label{eq:Heff}
\end{equation}
where $m_\alpha$ is the mass of the active neutrino and $m_s$ the mass of the sterile neutrino. Furthermore, the potential of interaction in the flavor basis is given by
\begin{equation}
V_{int}=\left(\begin{array}{cc} 
V_{\nu_\alpha \chi} & 
0 \\ 
& \\ 
0 & 
V_{\nu_s \chi} \\
 \end{array} \right)\, ,
\label{eq:Veff}
\end{equation}
The potential $V_{\nu_\alpha \chi}$ takes into account a possible interaction between active neutrinos and dark matter via a boson with light masses as for instance the one studied in \cite{vandenAarssen:2012vpm} and it can be computed as \cite{Miranda:2013wla}
\begin{equation}
V_{\nu_\alpha \chi} \sim  \frac{g_{\nu_\alpha} g_\chi}{m_I^2} N_\chi =\varepsilon_{{\nu_\alpha}\chi}G_FN_\chi \,
\end{equation}
where $g_{\nu_\alpha}, g_\chi$ are the coupling constants, $m_I$ the mass of the intermediate boson and $N_\chi$ the average number density of dark matter particles which for the case of the Milky Way is of the order $N_\chi=3\times10^{-16}$ eV$^3$.
Furthermore we have included the effect of a potential for the interaction of sterile neutrinos with dark matter $V_{\nu_s \chi}$. This interaction appears in different extensions of
the Standard Model, where many dark particles, including sterile
neutrinos, could populate the dark sector and interact among
themselves~\cite{Berezhiani:1995yi,Laha:2013xua}.
\begin{equation}
V_{\nu_s \chi} \sim  \frac{g_{\nu_s} g_\chi}{m_I^2} N_\chi =\varepsilon_{{\nu_s}\chi}G_FN_\chi\,.
\end{equation}
The parameters $\varepsilon_{\nu_{\alpha}\chi}$ and $\varepsilon_{s\chi}$ account for the coupling strength in terms of the Fermi constant $G_F$.
We have already neglected the well-known interaction potential of the active neutrino with ordinary
fermions since it is negligible compared with $V_{\nu_s \chi}$ and $V_{\nu_\alpha \chi}$.The interaction potential  $V_{\nu_s f}$ has already been studied~\cite{Bramante:2011uu} and is negligible compared to the other potentials. Therefore, it will be neglected as well.
Finally, 
 \begin{equation}
U=\left(\begin{array}{cc} 
\cos\theta_0 & 
-\sin\theta_0 \\ 
& \\ 
\sin\theta_0 & 
\cos\theta_0 \\
 \end{array} \right)\, ,
\label{eq:rotacion}
\end{equation}
with the angle $\theta_{0}$ is the vacuum mixing angle between the
sterile  and  the active neutrino. 
Withe the inclusion of the modified energy dispersion eq. \ref{dispersion}, we can diagonalize $\rm{H}_{eff}$ and compute the survival oscillation probability with matter and a generalized uncertainty principle which is given by

\begin{equation}
    P_{\nu_{\alpha} \rightarrow \nu_s}=\frac{\sin^{2}(2\theta_{0})\sin^{2}(\pi \frac{L_{0}^{osc}}{L_{m}^{osc}})}{\left(\cos(2\theta_0)-\frac{2E\sqrt{1-2\beta E^2}}{\Delta m^2}|\epsilon|G_F N_\chi\right)^{2}+\sin^{2}(2\theta_{0})}\label{probabilidad}
\end{equation}    
With
\begin{equation}
L_{m}^{osc}=\frac{L_{0}^{osc}}{\sqrt{\left(\cos(2\theta_0)-\frac{2E\sqrt{1-2\beta E^2}}{\Delta m^2}|\epsilon|G_F N_\chi\right)^{2}+\sin^{2}(2\theta_{0})}}\label{distancia}
\end{equation}
and $L_0^{osc}=\frac{4E}{\Delta m^2}$. Furthermore $|\epsilon|=\varepsilon_{\nu_{\alpha}\chi}-\varepsilon_{s\chi}$ and we have defined $\Delta m^2=m_\alpha^2-m_s^2$ the mass square difference between the active and the sterile neutrino.

\section{Resonant effects with the inclusion of a generalized uncertainty principle}\label{seccion3}
From Eq. \ref{probabilidad}, it is clear that there is a resonance when
\begin{equation}
\Delta m^{2}\cos2\theta_{0} = 2E\sqrt{1-2\beta E^2}G_F|\epsilon|N_\chi] \,.
\label{resonance}
\end{equation}
\subsection{No GUP: Case with $\beta=0$}
For $\beta=0$, with a constant density of dark matter $N_\chi=\rho_{DM}/m_\chi$, there is a resonance at 
energy given by
\begin{equation}
E_R=\frac{\Delta m^2 \cos(2\theta_0)}{2 G_F |\epsilon|N_\chi}\,. \label{energiaR}
\end{equation}
Motivated by the possible deficit in the UHE neutrino flux in the energy window $2\times10^{14}\rm{eV}<E<1\rm{PeV}$,
in \cite{Miranda:2013wla} it was found that for a mass squared difference between the active and the sterile neutrino $\Delta m^2=10^{-13}\rm{eV}^2$, a dark matter particle mass of $m_\chi=2\times 10^{10}\rm{eV}$, a coupling  $|\epsilon|=3\times 10^{11}$ motivated by the possible interaction of neutrinos with dark matter with a light intermediate boson \cite{vandenAarssen:2012vpm} and $\theta_0=\pi/12$, then $E_R=1\times 10^{14} \rm{eV}$. Thus active UHE neutrinos will resonant convert to sterile neutrinos and it could explain the observed deficit at Ice-Cube detector. 

\begin{figure}
    \centering
    \includegraphics[width=0.9\linewidth]{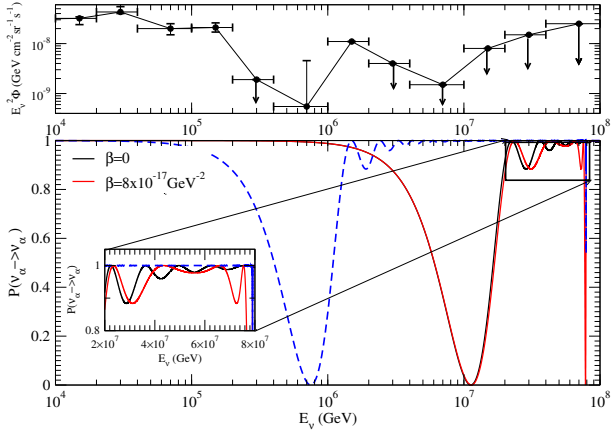}
    \caption{Survival probability as a function of the neutrino energy. It is observed that there is resonance at the energy window $2\times10^{14}\rm{eV}<E<1\rm{PeV}$.}
    \label{fig:seconddip}
\end{figure}
\subsection{The inclusion of GUP in neutrino oscillations: the case $\beta\ne0$}
GUP induces a new dispersion relation (see Eq. \ref{dispersion} that modifies the probability of neutrino oscillations in constant matter like it is expressed in Eq. \ref{probabilidad}. This modified probability has an slightly different resonance condition. Indeed, for $\beta \ne 0$, there are two energies where the resonant conversion takes place. 
Eq. \ref{resonance} can be written with the help of $E_R$ given by Eq. \ref{energiaR} as
\begin{equation}
    E\sqrt{1-2\beta E^2}=E_R\,,
\end{equation}
that has two solutions: 
\begin{equation}
    E_{NR}^2=\frac{1 \pm \sqrt{1-8\beta E_R^2}}{4\beta}\,.\label{nuevas}
\end{equation}
The parameter $\beta$ that appears in the GUP is expected to be vary small. 
In particular, if $8\beta E_R^2\ll1$ we can approximate Eq. \ref{nuevas} as
\begin{equation}
 E_{NR}^2\sim\frac{1}{4\beta}\left(1 \pm (1-4\beta E_R^2)\right)\,,
 \end{equation}
or, in other words, we have one resonance at $E_{NR}\simeq E_R$ and a new second energy where there is a resonant conversion at energy
\begin{equation}
E_{NR}^2\simeq\frac{1}{2\beta}-E_R^2\,.
\end{equation}
In Figure \ref{fig:seconddip} we show the survival probability for the same parameters as the case discussed in \cite{Miranda:2013wla} with the addition of the GUP parameter $\beta=8\times10^{-17}\rm{GeV}^{-2}$.
It is clear that there is a new resonance located at $E_{NR}$. 
\section{Conclusions}\label{seccion4}
Neutrinos are perhaps the less understood particle within the standard model. Nevertheless, it is for sure the best particle to explore the universe at the highest energies. In particular, UHE neutrinos could play an invaluable role in the search for a theory of quantum gravity. 
In this note we have complemented the scenario of a resonant neutrino conversion into sterile neutrinos enhanced by the interaction of neutrinos with dark matter \cite{Miranda:2013wla} by incorporating a generalized uncertainty principle.
We have shown that GUP induces a second resonance in the neutrino oscillation probability. This second resonance will be a clear signature of GUP that will be measurable as a deficit in the UHE neutrino flux.

%{References}
\bibliographystyle{unsrt}
\bibliography{referencias.bib} 
%%%%%%%%%%%%%%%%
\end{document}